\documentclass[aps,prd,reprint,preprintnumbers,superscriptaddress,showpacs,twocolumn]{revtex4-1}
\usepackage{latexsym,graphicx,amssymb,amsmath,mathrsfs}
\usepackage{setspace,bm}
\usepackage[breaklinks, colorlinks=true, pdfstartview=FitV, linkcolor=red, citecolor=blue, urlcolor=blue]{hyperref}
\usepackage{amsmath}
\usepackage[usenames]{color}
\usepackage{latexsym}
\usepackage{epstopdf}
\usepackage{amssymb}

\newcommand{\Slash}[1]{{\ooalign{\hfil/\hfil\crcr$#1$}}}
\newcommand\nc{N_\mathrm{c}}
\newcommand\nf{N_\mathrm{f}}

\usepackage[normalem]{ulem}  % \sout{old text} for strikeout

\newcommand{\comment}[1]{}

\renewcommand\sout{\bgroup \color{red} \ULdepth=-.5ex \ULset}
\begin{document}
\preprint{YITP-18-43, KUNS-2724}

\title{ Control the model sign problem via path optimization method:\\
Monte-Carlo approach to QCD effective model with Polyakov loop
}

\author{Kouji Kashiwa}
\email[]{kouji.kashiwa@yukawa.kyoto-u.ac.jp}
\affiliation{Fukuoka Institute of Technology, Wajiro, Fukuoka 811-0295,
Japan}
\affiliation{Yukawa Institute for Theoretical Physics,
Kyoto University, Kyoto 606-8502, Japan}

\author{Yuto Mori}
\email[]{mori@ruby.scphys.kyoto-u.ac.jp}
\affiliation{Department of Physics, Faculty of Science, Kyoto
University, Kyoto 606-8502, Japan}

\author{Akira Ohnishi}
\email[]{ohnishi@yukawa.kyoto-u.ac.jp}
\affiliation{Yukawa Institute for Theoretical Physics,
Kyoto University, Kyoto 606-8502, Japan}

\begin{abstract}
We apply the path optimization method to a QCD effective model with the
 Polyakov loop at finite density to circumvent
 the model sign problem.
The Polyakov-loop extended Nambu--Jona-Lasinio model is employed as the
 typical QCD effective model and then the hybrid Monte-Carlo method is
 used to perform the path integration.
To control the sign problem, the path optimization method is used with
 complexification of temporal gluon fields to modify the integral path
 in the complex space.
We show that the average phase factor is well improved on the modified
 integral-path compared with that on the original one.
This indicates that the complexification of temporal gluon fields may be
 enough to control the sign problem of QCD in the path optimization method.
 \end{abstract}

%\pacs{11.30.Rd, 21.65.Qr, 25.75.Nq}
\maketitle

\section{Introduction}

Investigation of the phase structure of Quantum Chromodynamics (QCD)
is one of the important subjects in the study of the hot and dense QCD
matter.
If we directly obtain the phase diagram at finite temperature ($T$) and
density from the first-principles calculation
such as the lattice QCD simulation, there are no fog in
the exploration.
Lattice QCD, however, has the sign problem at finite real chemical
potential ($\mu$) and then we cannot obtain the reliable
results at finite density.
To circumvent the sign problem, several methods have been proposed such
as the Taylor expansion method~\cite{Miyamura:2002en,Allton:2005gk,Gavai:2008zr},
the reweighting
method~\cite{Fodor:2001au,*Fodor:2001pe,*Fodor:2004nz,Fodor:2002km},
the analytic continuation
method~\cite{deForcrand:2002ci,*deForcrand:2003hx,D'Elia:2002gd,*D'Elia:2004at,Chen:2004tb},
the canonical
approach~\cite{Hasenfratz:1991ax,Alexandru:2005ix,Kratochvila:2006jx,deForcrand:2006ec,Li:2010qf}
and so on.
However, we cannot go beyond the $\mu/T \sim 1$ line by using those
methods; for example, see Ref.~\cite{deForcrand:2010ys}.

Recently, new ideas have been applied to attack the sign problem such as
the complex Langevin method~\cite{Parisi:1980ys,Parisi:1984cs} and the
Lefschetz-thimble method~\cite{Witten:2010cx,Cristoforetti:2012su,Fujii:2013sra}.
Both methods are based on the complexification of dynamical variables.
The complex Langevin method is based on the stochastic quantization
and thus it does not have the sign problem, in principle.
In the Lefschetz-thimble method, one should solve flow equations to
construct the new integral path which is corresponding to the steepest
descent trajectory; this trajectory is so called the Lefschetz thimble.
Unfortunately, these methods have some serious problems and thus it is
difficult to apply it to QCD at high density:
In the complex Langevin method, there is the possibility that
it is  converged to wrong results due to the excursion and singular
problems~\cite{Aarts:2009uq,Tanizaki:2015pua}.
In comparison, the Lefschetz-thimble method has the global sign problem when
multi Lefschetz-thimbles contribute to the path integral and then there is the
serious cancellation between them; for example, see
Ref.~\cite{Fujii:2013sra}.
In addition, we should evaluate the Jacobian induced by the modification of the
integral path and it leads the serious increase of the numerical cost.
Furthermore, the Jacobian induces the residual sign problem because
the oscillation of the Boltzmann weight arises.
Also, we may face the problem to draw thimbles when the classical or the
effective action has singular points~\cite{Mori:2017zyl}.

In Ref.~\cite{Mori:2017pne}, we have proposed the new method which we call the
{\it path optimization method} (POM) motivated by the Lefschetz-thimble
method.
The path optimization method is strongly improved in
Ref.~\cite{Mori:2017nwj} by introducing the feedforward neural network
to optimize the modified integral path.
In this method, we first complexify the variables of integration as
in the Lefschetz-thimble method.
Then, the modified integral path is
constructed to minimize
the cost function which reflects the seriousness of the sign problem.
Therefore, we can treat the sign problem as the optimization problem.
The sign problem in the simple one-dimensional
integration~\cite{Mori:2017pne,Ohnishi:2017zxh} and
the complex $\lambda \phi^4$ theory~\cite{Mori:2017nwj}
are found to be under controlled.
Unfortunately, the numerical cost of the path optimization method is
still heavy and thus we cannot apply it to QCD yet, but this method has
large extensibility compared with the Lefschetz-thimble
method and thus we can still dream to apply it to
QCD in the future.

Because of several difficulties in QCD as mentioned above, QCD effective
models have been widely used to investigate the QCD phase structure
because such effective model is much easier than original
QCD.
The Polyakov-loop extended Nambu--Jona-Lasinio (PNJL)
model~\cite{Meisinger:2001cq,Fukushima:2003fw,Ratti:2005jh} is one of the
famous and powerful effective models.
Unfortunately, the PNJL model has the model sign
problem~\cite{Fukushima:2006uv,Tanizaki:2015pua} even in the
mean-field approximation because one particular global minimum perfectly
dominate the path integration and then the thermodynamic potential
can be complex.
It should be noted that this complex nature of the thermodynamic
potential is not related with instabilities; the correct thermodynamic
potential should be real.

In this study, we consider the PNJL model and the path optimization method
to circumvent the model sign problem.
This article is organized as follows.
In Sec.~\ref{Sec:PNJL} and \ref{Sec:POM}, we explain the formulation of
the PNJL model and the path optimization method, respectively.
The numerical results are shown in Sec.~\ref{Sec:Numerical}.
Section \ref{Sec:Summary} is devoted to summary.

\section{Formulation}
\label{Sec:PNJL}

In this work, we employ the PNJL model as the QCD effective model.
The PNJL model can describe the chiral symmetry breaking/restoration and
the approximate confinement-deconfinement transition.
Also, it can well reproduce QCD properties at finite imaginary chemical
potential which is a big advantage from the viewpoint of the topologically
determined confinement-deconfinement
transition~\cite{Kashiwa:2015tna,Kashiwa:2016vrl,Kashiwa:2017yvy}.
The following formulation and the computation of the Monte-Carlo PNJL
model is based on Ref.~\cite{Cristoforetti:2010sn}.

\subsection{Polyakov-loop extended Nambu--Jona-Lasinio model}
The Euclidean action of the PNJL model is
\begin{align}
 \Gamma_\mathrm{PNJL}
 &= \int d^4 x_\mathrm{E} \Bigl[
    {\bar q} (-i\Slash{D} + m_0 - \mu \gamma_0) q
 \nonumber\\
          &- G \Bigl\{ ({\bar q}q)^2
           + ({\bar q} i {\vec \tau} \gamma_5 q)^2 \Bigr\} \Bigr]
- \beta V {\cal V}_{\Phi} (\Phi,{\bar \Phi}),
\end{align}
where $\beta = 1/T$, $V$ is the three-dimensional spatial volume,
$q$ denotes the two-flavor quark-fields,
$m_0$ does the current quark mass,
$D_\nu = \partial_\nu  - i A_\nu \delta_{\nu 4}$,
$G$ is the coupling constant of the four-fermi interaction,
$\Phi$ (${\bar \Phi}$) means the Polyakov loop (its conjugate) and
${\cal V}_{\Phi}$ expresses the gluonic contribution.

With the homogeneous auxiliary-field ansatz after the
Hubbard-Stratonovich transformation, the effective action is
simplified as $\Gamma = \beta V {\cal V}$.
Then, the effective potential is expressed as
\begin{align}
{\cal V} &= {\cal V}_\mathrm{NJL} + {\cal V}_{\Phi},
\end{align}
where ${\cal V}_\mathrm{NJL}$ is the contributions of the NJL part.
In the actual calculation, we employ the Polyakov gauge, $\partial_4
A_4=0$.

The actual form of ${\cal V}_\mathrm{NJL}$ becomes
\begin{align}
 {\cal V}_\mathrm{NJL}
 &= - 2 \nf \int_\Lambda \frac{d^3 p}{(2\pi)^3}
      \Bigl[ \nc E_{\bf p} - \nc \sqrt{{\bf p}^2 + m_0^2}
\nonumber\\
         &+ T \ln \Bigl( f^- f^+ \Bigr)\Bigr]
  + G(\sigma^2 + {\vec \pi}^2),
\end{align}
where $\nf$ ($\nc=3$) is the number of flavor (color),
$\Lambda$ is the three-dimensional momentum cutoff and
\begin{align}
f^-
&= 1
 + 3 (\Phi+{\bar \Phi} e^{-\beta E_{\bf p}^-} ) e^{-\beta E_{\bf p}^-}
 + e^{-3\beta E_{\bf p}^-},
\nonumber\\
f^+
&= 1
 + 3 ({\bar \Phi}+\Phi e^{-\beta E_{\bf p}^+} ) e^{-\beta E_{\bf p}^+}
 + e^{-3\beta E_{\bf p}^+},
\end{align}
with $E_{\bf p}^\mp = E_{\bf p} \mp \mu
                    = \sqrt{\varepsilon_{\bf p}^2 + 2 N^+N^-} \mp \mu$
and $\varepsilon_{\bf p} = \sqrt{{\bf p}^2 + M^2+ N^2}$.
The auxiliary fields are redefined as
\begin{align}
 M = m_0 - 2 G \sigma,~~~N = - 2 G \pi^0,~~~N^\pm = - 2 G \pi^\pm.
\end{align}
with $\pi^0 = \pi_3$ and $\pi^\pm = (\pi_1 \pm i \pi_2)/ \sqrt{2}$.
The functional form of the Polyakov loop and its conjugate are
\begin{align}
\Phi
&= \frac{1}{\nc} \mathrm{tr}_c
   \Bigl[ e^{i \beta A_4} \Bigr]
 = \frac{1}{\nc}
   \Bigl[ e^{i \beta \phi_1} + e^{i \beta \phi_2} + e^{i \beta \phi_3}
   \Bigr],
\nonumber\\
{\bar \Phi}
&= \frac{1}{\nc}
   \Bigl[ e^{-i \beta \phi_1}
        + e^{-i \beta \phi_2}
        + e^{-i \beta\phi_3}
   \Bigr],
 \label{Eq:Pol}
\end{align}
where
\begin{align}
 \phi_1 &= A_3 + \frac{1}{\sqrt{3}}A_8,~~~
 \phi_2  = - A_3 + \frac{1}{\sqrt{3}}A_8
 \nonumber\\
 \phi_3 &= - (\phi_1 + \phi_2) = - \frac{2}{\sqrt{3}}A_8,
\end{align}
and then $A_4$ are diagonalized because we use the Polyakov gauge;
$A_4 =
\mathrm{diag}(A_3+A_8/\sqrt{3},-A_3+A_8/\sqrt{3},-2A_8/\sqrt{3})$.

In this paper, we choose the logarithmic Polyakov-loop
potential~\cite{Roessner:2006xn} as the gluonic contribution;
\begin{align}
\frac{{\cal V}_{\Phi}}{T^4}
 &= - \frac{1}{2} a_T {\bar \Phi} \Phi + b_T \ln (h),
\end{align}
where
\begin{align}
h &= 1 - 6 {\bar \Phi} \Phi
       + 4 ({\bar \Phi}^3 + \Phi^3)
       - 3 ({\bar \Phi} \Phi)^2,
\end{align}
with
\begin{align}
 a_T &= a_0 + a_1 \Bigl(\frac{T_0}{T} \Bigr)
            + a_2 \Bigl(\frac{T_0}{T} \Bigr)^2,~~~
 b_T  = b_3 \Bigl(\frac{T_0}{T} \Bigr)^3.
\end{align}
The parameters should be set to reproduce the lattice QCD data in the pure
gauge limit.

\subsection{System volume and parameters}
In this study, we follow the lattice formalism and thus
$V$ can be expressed~\cite{Cristoforetti:2010sn} as
\begin{align}
 V = N_s^3 a^3 = \frac{N_s^3}{T^3 N_t} = \frac{k}{T^3},
\end{align}
where
$N_s$ ($N_t$) are the number of spatial (temporal) lattice sites and
$a$ is the lattice spacing.
Then, $V$ should depend on the temperature.
In this article, we use the homogeneous ansatz and thus our
Monte-Carlo simulation reaches the mean-field results in the
infinite volume limit.
This fixed $k$ treatment with homogeneous
ansatz leads the inconsistent results, in principle.
However, such inconsistency becomes smaller and smaller when the system
volume becomes larger and larger and thus it is a minor problem in this
study.
Full simulation of the lattice PNJL model is our future work.

The present PNJL model has three parameters, $G$, $m_0$ and $\Lambda$ in
the NJL part.
The actual values of the parameters are taken from
Ref.~\cite{Kashiwa:2006rc};
$m_0=5.5$ MeV, $G=5.498$ GeV$^{-2}$ and $\Lambda=631.5$ MeV.
The parameters in the Polyakov-loop potential is
taken from Ref.~\cite{Roessner:2006xn};
\begin{align}
 a_0&=3.51,~~a_1=-2.47,~~a_2=15.2,~~b_3=-1.75,
\end{align}
with $T_0 = 270$ MeV.

\section{Path optimization method}
\label{Sec:POM}

To deal with the model sign problem appearing at finite density,
we here introduce the path optimization method.

\subsection{Introduction to POM}
In the path optimization method, we start from the complexification of
the variables of the integration,
$x_i \in \mathbb{R} \to z_i \in \mathbb{C}$
where $i=1, \cdots, n$ with the dimension of integration $n$.
To construct the new integral path in the complex plane, we prepare the
cost function which should be related with the seriousness of the sign
problem.
The functional form of the new integral path is represented by
using the feedforward neural network with some parameters which are
optimized via the minimization of the cost function.
Since the neural network even with the mono hidden-layer can
approximate any kind of continuous function on the compact subset as
long as we can use sufficient number of units in the
layer~\cite{cybenko1989approximation,hornik1991approximation}, the
neural network seems to be suitable for the path optimization method.
Details are shown in Ref.~\cite{Mori:2017pne}.

\begin{figure}[t]%[H]
%%\begin{figure}[!t]%[H]
 \centering
 \includegraphics[width=0.35\textwidth]{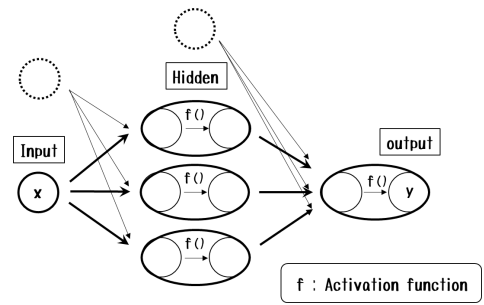}
 \caption{ The schematic figure of the feedforward neural network used
 in this study for the case with the single input and output.
 The dotted circles mean the bias and the thick arrows indicate the
 weight. In the final step, we calculate Eq.~(\ref{Eq:fo}) to obtain the
 modified integral-path.
 }
\label{Fig:ml}
\end{figure}

In the path optimization method, we represent $z_i$ by using the
parametric quantity ($t$) as
\begin{align}
 z_i(t) &= t_i + i [ w_i f_i(t) + b_i ],
 \label{Eq:fo}
\end{align}
where $w$ and $b$ are parameters.
Parameters $w$, $b$ and also parameters in $f$ are determined by using
the back-propagation algorithm~\cite{rumelhart1988learning}.
Figure \ref{Fig:ml} shows the schematic figure of the neural
network used in this study.
%%%%%%%%%%%%%%%%%%%% Fig %%%%%%%%%%%%%%%%%%%%%%%%
%%%%%%%%%%%%%%%%%%%%%%%%%%%%%%%%%%%%%%%%%%%%%%%%%
In the back-propagation, we choose $\tanh$ for the activation function.
In the following, we represents parameters in the neural network as $c=\{c_i\}$.
Details are shown in Refs.~\cite{Mori:2017pne,Mori:2017nwj}.
It should be noted that the path optimization method leads the same results
with the original theory because of the Cauchy(-Poincare)'s theorem as
long as the integral path does not go across singular points,
$Je^{-\Gamma} = 0$.
In the complex Langevin method, singular points induce the problem
because it leads the singular drift term in the Langevin-time evolution,
the path optimization method can care for it.

In this study, we complexify $A_3$ and $A_8$ and then $\sigma$ and
${\vec \pi}$ are still treated as real variables.
Since it is known that the model sign problem can be resolved by the
complexification of the temporal gluon fields in the Lefschetz-thimble
method~\cite{Tanizaki:2015pua}, our treatment should work.
In addition, such treatment is consistent with the calculation which
imposes the ${\cal C} {\cal K}$ symmetry on the fermion determinant at
finite density~\cite{Nishimura:2014rxa,Nishimura:2014kla}
where ${\cal C}$ and ${\cal K}$ denote the charge conjugation and the
complex conjugation, respectively.
Unfortunately, the Lefschetz-thimble method is difficult to apply to the
PNJL model because we should solve flow equations and it
sometimes encounter singular pints of the effective
potential.
The NJL model with the repulsive vector-type interaction is a
typical example~\cite{Mori:2017zyl}.
In comparison, the path optimization method can avoid the problem and
thus it is suitable for the PNJL model analysis.

It should be noted that the present path optimization with the machine
learning is unsupervised learning because we do not need teacher data
to obtain optimized parameters in the neural network.
We try to increase the average phase factor compared with that in the
previous optimization step.
The one attempt to introduce the supervised learning to the study of the
sign problem has been done in Ref~\cite{Alexandru:2017czx} based on
the generalized Lefschetz-thimble method~\cite{Alexandru:2015sua}.

\subsection{Optimization process}
We use the following cost function in the calculation;
\begin{align}
 {\cal F}[z(t)]
 &= \frac{1}{2} \int d^nt~
   |e^{i \theta(t)} - e^{i \theta _0}|^2 \times |J(t) e^{-\Gamma(z(t))}| \nonumber \\
 &= \int d^nt~|J(t) e^{-\Gamma(z(t))}|
  \nonumber\\
 & \hspace{1.5cm}
    - \left |\int d^nt~J(t) e^{-\Gamma(z(t))} \right |,
\label{Eq:cf}
\end{align}
where
\begin{align}
 \theta(t) &= \arg (J(t)e^{-\Gamma(z(t))}), ~~~
 \theta_0   = \arg ({\cal Z}),
 \nonumber\\
 J(t) &= \det \Bigl( \frac{\partial z_i}{\partial t_j} \Bigr),
\end{align}
with the partition function ${\cal Z}$.
In the equation, $z_i$ represent complexfied $A_3$ and $A_8$ and also the real $\sigma$
and ${\vec \pi}$. Thus, we have eight dynamical variables.
If we wish to take care of the periodicity of the effective potential
for $\mathrm{Re}~A_3$ and $\mathrm{Re}~A_8$, we should use periodic
functional form for those as inputs of
the neural network like as Ref.~\cite{Alexandru:2017czx}.
In this study, configurations are well localized in the range
$-\pi \le \mathrm{Re}~A_3/T \le \pi$ and
$-\pi \le \mathrm{Re}~A_8/(T\sqrt{3}) \le \pi$ by using
the simple complexification of $A_3$ and $A_8$.
Thus, we do not introduce the periodic form of inputs in this study.
This cost function is proportional to
$\langle e^{i\theta} \rangle$ and thus
it reflects the seriousness of the sign problem~\cite{Mori:2017nwj}.
In the physical system, $\theta_0$ should be $0$.
If we consider $\theta_0 \neq 0$,
we can apply the path optimization method to other systems
such as the complex chemical potential.

In the hybrid Monte-Carlo method, we should evaluate the expectation
value of the cost function.
To do this, we replace the cost function as
\begin{align}
 {\cal F}[z(t)] \to  \frac{{\cal F}[z(t)]}
                          {\int d^n t~{\cal P}(t)},
\label{Eq:F}
\end{align}
where ${\cal P}$ is the appropriate probability distribution.

In the back-propagation procedure, we need the derivative of the cost
function by $c_i$. We represent it as $d {\cal F}_i$ below.
After straightforward calculations, we finally reach the expression;
\begin{align}
d {\cal F}_i (t, c) &=
   |J(t)e^{-\Gamma(z(t))}| \nonumber \\
 & \hspace{-1.0cm}
 \times {\rm Re} \left[ (1-e^{i(\theta(t) - \theta_0)})
 \frac{\partial}{\partial c_i} \log(J(t)e^{-\Gamma(z(t))}) \right],
\label{Eq:dF}
\end{align}
In the hybrid Monte-Carlo method, we rewrite Eq.~(\ref{Eq:dF}) as
similar to Eq.~(\ref{Eq:F}) with ${\cal P}$.
This cost function is responsible to the alignment of the Boltzmann weight
with each other on the modified integral path if the
points are relevant to the path-integral.

To make our optimization easier, we employ the mini-batch training.
The configurations are divided as $N_\mathrm{config} = n N_\mathrm{batch}$
where $N_\mathrm{batch}$ is the batch size and then the learning is
performed batch by batch.
To include all updation of each batch,
the parameters in the feedforward neural network is updated by
replacing $dF_i$ by its mean-value as
\begin{align}
d {\cal F}_i(t,c) \to
\frac{1}{N_\mathrm{batch}} \sum _{k=1}^{N_\mathrm{batch}} d{\cal F}^{(k)}(t, c).
\end{align}
In one optimization step, we update $n$-times
with $N_{\mathrm{batch}}$ configurations.

In this study, we use the simple feedforward neural network with the
hidden mono-layer.
Input is the original integral path and output is its imaginary part.
For the optimizer, we employ Adam algorithm~\cite{kingma2014adam};
\begin{align}
v_i^{(j+1)} &= \beta_1 v^{(j)}  + (1-\beta_1) d{\cal F}_i,\nonumber\\
r_i^{(j+1)} &= \beta_2 r^{(j)} + (1 - \beta_2) d{\cal F}_i^2,\nonumber\\
{\hat v}_i^{(j+1)} &= \frac{v^{(j+1)}_i}{1-\beta_1^j},~~~
{\hat r}_i^{(j+1)}  = \frac{r^{(j+1)}_i}{1-\beta_2^j},\nonumber\\
c^{(j+1)}_i &= c^{(j)}_i
             - \frac{\eta}{\sqrt{{\hat r}^{(j+1)}_i}+\epsilon}
               {\hat v^{(j+1)}_i},
\end{align}
where $j$ is the fictitious time step, $d{\cal F}_i$ means
$\partial {\cal F}/{\partial c_i}$ and
$\beta_1$ and $\beta_2$ are the smoothing factors of the exponential
moving average.
This algorithm is based on the AdaGrad
algorithm~\cite{duchi2011adaptive} and
the momentum method with preventing the learning weight decay.

\subsection{Simulation setup}
The number of units in the hidden layer is set to $N_\mathrm{unit} = 4$.
For Adam algorithm,
we use $\eta=0.001$,
$\alpha = 0.999$, $\beta = 0.9$ and $\epsilon = 10^{-8}$.
In the mini-batch training, we set to $N_\mathrm{batch}=10$.
These parameters are so called hyper parameters in the machine learning.
Initial values of parameters in the neural network are prepared based on
Xavier initialization~\cite{glorot2010understanding}.

In the calculation of the expectation values of operators,
we have generated $80000$ configurations analyzed each $50$
trajectories for each optimization step.
Then, the expectation values are estimated after $2$
times optimization.
Statistic errors are obtained by using the Jack-knife method where the
bin number is set to $10$.
The expectation value of an operator (${\cal O}$) is obtained via the
phase reweighting as
\begin{align}
 \langle {\cal O} \rangle
 &= \frac{\displaystyle \int d^n t~{\cal O} e^{i\theta}|J(t) e^{-\Gamma(z(t))}|}
         {\displaystyle \int d^n t~e^{i\theta}|J(t) e^{-\Gamma(z(t))}| }.
\end{align}
In this study, we calculate the chiral condensate and the Polyakov loop.

\section{Numerical results}
\label{Sec:Numerical}

\begin{figure}[t]%[H]
%%\begin{figure}[!t]%[H]
 \centering
 \includegraphics[width=0.33\textwidth]{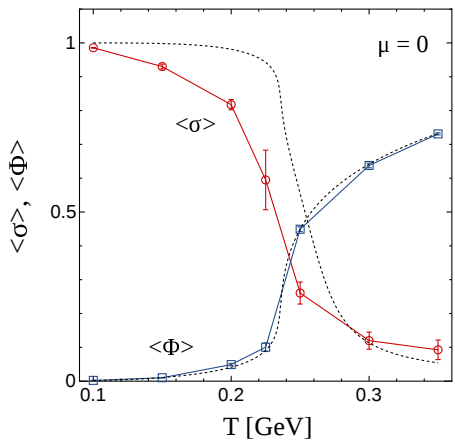}
 \caption{The $T$-dependence of $\langle \sigma \rangle $ and
 $\langle \Phi \rangle$
 at $\mu=0$ where $\langle \sigma \rangle$ is normalized by that at
 $T=\mu=0$ in the infinite volume limit.
 The dotted lines are mean-field results in the infinite volume limit as
 the eye guide.
 }
\label{Fig:T-dep}
\end{figure}

The $T$-dependence of the chiral condensate and the Polyakov loop at $\mu=0$
is shown in Fig.~\ref{Fig:T-dep}.
The mean-field results in the infinite volume limit is also shown as the
eye guide.
%%%%%%%%%%%%%%%%%%%% Fig %%%%%%%%%%%%%%%%%%%%%%%%
%%%%%%%%%%%%%%%%%%%%%%%%%%%%%%%%%%%%%%%%%%%%%%%%%
Because of the finite size effect, $\langle \sigma \rangle$ is deviated
from the mean-field results in the infinite volume limit and
this result is consistent with results obtained in
Ref~\cite{Cristoforetti:2010sn}.
In the present calculation, $V$ depends
on $T$ and thus the finite size effect becomes serious
when $T$ increases.
The expectation values of ${\vec \pi}$ are
consistent with zero in
$2\sigma$ error.

\begin{figure}[t]%[H]
%%\begin{figure}[!t]%[H]
 \centering
 \includegraphics[width=0.4\textwidth]{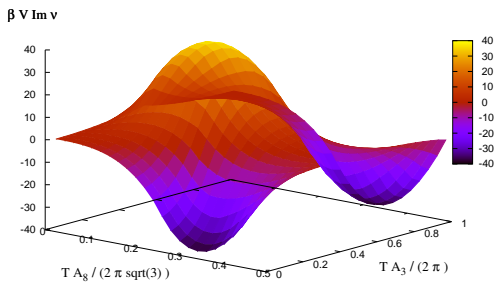}
 \caption{ The imaginary part of the effective potential in the
 $A_3$-$A_8$ plane with $\sigma^2 = {\vec \pi}^2=0$.
 The effective action is normalized as $\beta V {\cal V}$ with $k=64$.
 }
\label{Fig:Im}
\end{figure}

To discuss the model sign problem, we consider the finite $\mu$ below.
Figure \ref{Fig:Im} shows the imaginary part of the effective action.
%%%%%%%%%%%%%%%%%%%% Fig %%%%%%%%%%%%%%%%%%%%%%%%
%%%%%%%%%%%%%%%%%%%%%%%%%%%%%%%%%%%%%%%%%%%%%%%%%
We can clearly see that the effective potential can become complex if we
pick up a certain point as the solution of the mean-field approximation.

\begin{figure}[t]%[H]
%%\begin{figure}[!t]%[H]
 \centering
 \includegraphics[width=0.33\textwidth]{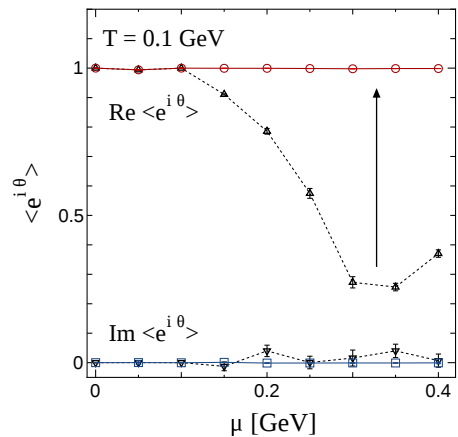}
 \caption{The $\mu$-dependence of $\langle e^{i\theta} \rangle$ at
 $T=100$ and $\mu=300$ MeV with $k=64$.
 The dotted and solid lines indicate results on the original
 and modified integral-paths, respectively.
 The modified integral-path is obtained after twice optimization.
 }
\label{Fig:APF-each}
\end{figure}

The $\mu$-dependence of the average phase factor on the original
integral-path at $T=100$ and $\mu=300$ is shown in Fig.~\ref{Fig:APF-each}.
%%%%%%%%%%%%%%%%%%%% Fig %%%%%%%%%%%%%%%%%%%%%%%%
%%%%%%%%%%%%%%%%%%%%%%%%%%%%%%%%%%%%%%%%%%%%%%%%%
We also show the results after twice optimization in the figure.
It can be clearly seen that $\langle e^{i\theta} \rangle$ becomes small
around the chiral transition region, $T \sim 335$ MeV, and then the model sign
problem seriously appears.
In the present computation, we use $k=64$ and thus
$\langle e^{i\theta} \rangle$ does not become small so much, but it
will be worse when we consider larger $k$ because
$\langle e^{i\theta} \rangle$ is exponentially suppressed by $k$.

\begin{figure}[b]%[H]
%%\begin{figure}[!t]%[H]
 \centering
 \includegraphics[width=0.33\textwidth]{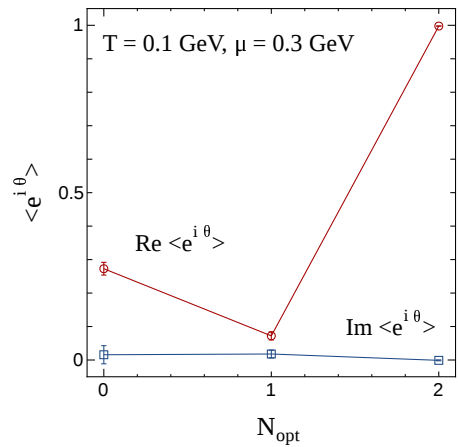}
 \caption{The average phase factor at each optimization step with
 $T=100$ and $\mu=300$ MeV.
 %The symbol $\mathrm{N_{opt}}$ means the number of the optimization.
 }
\label{Fig:apf}
\end{figure}

To improve the average phase factor, we use the path optimization method.
Figure \ref{Fig:apf} shows the average phase factor at each optimization
step with $T=100$ and $\mu = 300$ MeV as an example.
%%%%%%%%%%%%%%%%%%%% Fig %%%%%%%%%%%%%%%%%%%%%%%%
%%%%%%%%%%%%%%%%%%%%%%%%%%%%%%%%%%%%%%%%%%%%%%%%%
%%%%%%%%%%%%%%%%%%%% Fig %%%%%%%%%%%%%%%%%%%%%%%%
\begin{figure*}[t]%[H]
%%\begin{figure}[!t]%[H]
 \centering
 \includegraphics[width=0.33\textwidth]{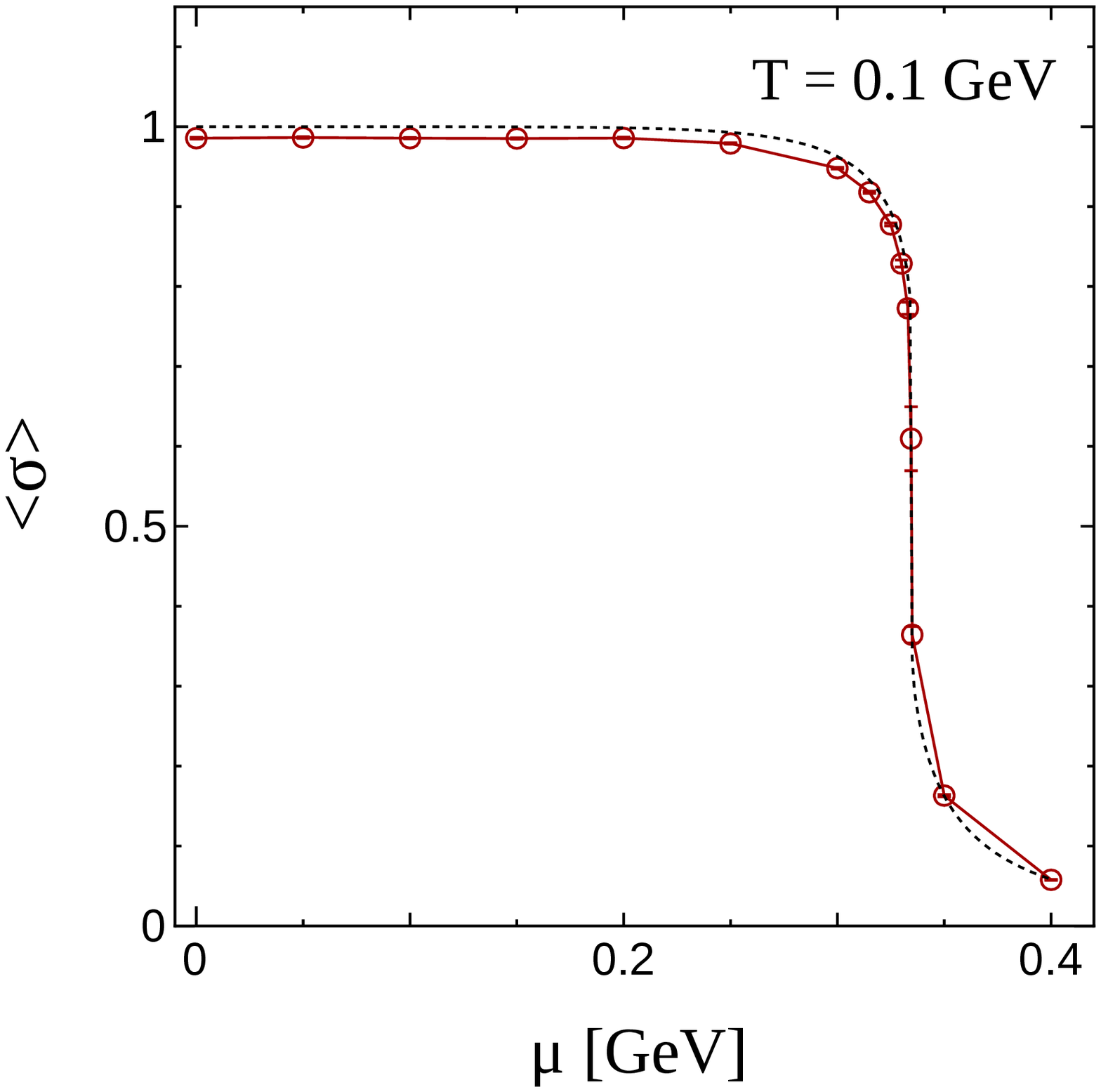}
 \hspace{10mm}
 \includegraphics[width=0.345\textwidth]{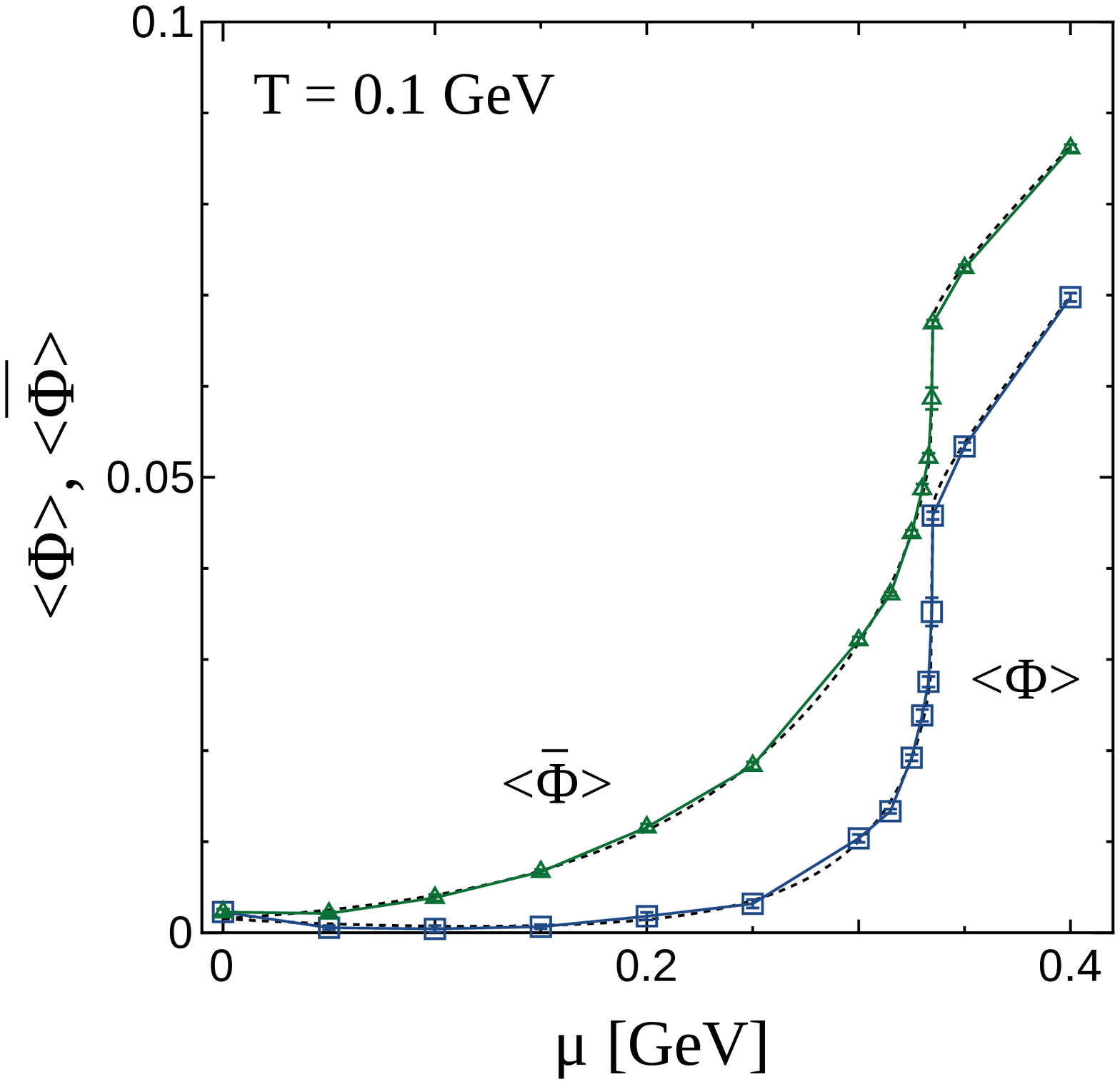}
 \caption{The left (right) panel shows the $\mu$-dependence of
 $\langle \sigma \rangle$
 ($\langle \Phi \rangle$) with $T=100$ and $\mu=300$ MeV where $\langle
 \sigma \rangle$
 is normalized by that at $T=\mu=0$ in the infinite volume limit.
 The dotted lines are mean-field results with imposing the ${\cal CK}$
 symmetry on the fermion determinant in the infinite volume limit as the
 eye guide.
 }
\label{Fig:OP}
\end{figure*}
%%%%%%%%%%%%%%%%%%%%%%%%%%%%%%%%%%%%%%%%%%%%%%%%%
After one optimization step ($\mathrm{N_{opt}}=1$), the average phase factor becomes worse.
Since the starting configurations used in the first optimization are
generated on the original integral-path and thus those are expected to be
located far from the relevant points of the optimized integral-path and
then the average phase factor becomes worse temporally.
However, the average phase factor is significantly improved after few
optimization steps.
In the case at $T=100$ and $\mu=300$ MeV with $k=64$,
$\mathrm{N_{opt}}=2$ are enough to make the average phase factor close to $1$.
It should be noted that the present tendency of the improvement depends
on several initial conditions of parameters and thus we should try with
several set of parameters when the optimization is not well worked.
Also, an increase of units in the hidden layer may lead to
better convergence~\cite{frankle2018lottery}.

The $\mu$-dependence of the chiral condensate and the Polyakov-loop
after performing the path optimization are shown in Fig.~\ref{Fig:OP}.
The mean-field results in the infinite volume with imposing the
${\cal C K}$ symmetry to the fermion determinant is also shown as the
eye guide.
We can correctly reproduce the relation
$\langle {\bar \Phi} \rangle \neq \langle \Phi \rangle$ with
$\langle {\bar \Phi} \rangle,$$\langle \Phi \rangle \in \mathbb{R}$
in the PNJL model at finite
$\mu$ by using the path optimization method.
These results mean that the path optimization method can well work in
the QCD effective model which has the model sign problem.
Also, we can expect that the complexification of the temporal gluon
field may be sufficient to control the sign problem in the lattice QCD with
the path optimization method since the lattice QCD and the PNJL model
share similar properties about the sign problem.

\section{Summary}
\label{Sec:Summary}

In this article, we apply the path optimization method to the
Polyakov-loop extended Nambu--Jona-Lasinio (PNJL) model to circumvent
the model sign problem.
This study is the first attempt to apply the path optimization method
to the effective model with dynamical quarks based on QCD.
The PNJL model can describe the chiral phase transition and also
approximated confinement-deconfinement transition and thus it is a good
starting point to investigate the QCD phase structure at finite real
chemical potential ($\mu$).
Therefore, we choose it as the typical QCD effective model in this
article.

The temporal components of the gluon field, $A_3$ and $A_8$, are
complexified
and the modified integral path in the complex space is expressed by
using the feedforward
neural network by minimizing the cost function which reflects the
seriousness of the sign problem.
Then, the sign problem becomes the optimization problem.
Parameters in the feedforward neural network
 are optimized via the back-propagation method.
The neural network tries to increase the average phase factor compared
with the previous optimization step and thus it is nothing but the
unsupervised learning.
The scalar and pseudo-scalar auxiliary fields are treated as real
valuables.
This treatment is motivated by the Lefschetz-thimble analysis done in
Ref.~\cite{Tanizaki:2015pua}.
In the actual optimization process, we use the mini-batch training with
Adam algorithm.

We have shown that our treatment of variables of integration works
well; the average phase factor is sufficiently improved after the
optimization at finite $\mu$.
This means that the path optimization method can resolve the model sign
problem based on the hybrid Monte-Carlo method.
In this study, we use the homogeneous ansatz for the auxiliary fields and
thus we cannot go beyond the mean-field approximation, but it is a
first step to correctly treat the sign problem in the QCD
effective models.
By considering the straightforward extension of the present formulation,
we will go beyond the mean-field approximation of the QCD effective
models. We leave the actual simulation as our future work.
From these results, we may expect that the complexification of the temporal gluon
fields is the sufficient way to control the sign problem in the lattice QCD with
the path optimization method
because QCD and the PNJL model share
similar properties about the sign problem.

\begin{acknowledgments}
This work is supported in part by the Grants-in-Aid for Scientific Research
 from JSPS (Nos. 15K05079, 15H03663, 16K05350),
the Grants-in-Aid for Scientific
 Research on Innovative Areas from MEXT (Nos. 24105001, 24105008),
 and by the Yukawa International Program for Quark-hadron
 Sciences (YIPQS).
\end{acknowledgments}

\bibliography{ref.bib}

\end{document}